\documentstyle[prl,aps]{revtex}
\begin{document}
\draft
\title{Brueckner approach to the spin-wave gap critical index for the two-layer Heisenberg antiferromagnet}
\author{P.V.Shevchenko$^{a}$, O.P.Sushkov}
\address
{School of Physics, The University of New South Wales,
Sydney 2052, Australia}
\date{28 September 1998}
\maketitle
\begin{abstract}
We consider the two-layer Heisenberg antiferromagnet 
near a zero temperature  quantum phase transition
from a disordered dimer phase to a collinear Neel state.
At approaching the transition point the spin-wave gap vanishes
as $\Delta \propto (J_\perp-J_{\perp c})^{\nu}$.
To account for strong correlations between the $S=1$ elementary excitations
we apply the Brueckner diagram approach which gives the critical index
$\nu\approx 0.5$. We demonstrate also that the linearized in density Brueckner
equations give the mean field result $\nu=1$. Finally an expansion of the 
Brueckner equations in powers of the density, combined with the scaling 
hypothesis, give $\nu\approx 0.67$. This value reasonably
agrees with that of the nonlinear O(3) $\sigma$-model.
Our approach demonstrates that for other quantum spin models the
critical index can be different from that in the
nonlinear $\sigma$-model. We discuss the conditions for this to occur.
\end{abstract}
\twocolumn
There is a significant interest in the physics of quantum phase transitions
in $S=1/2$ two-dimensional (2D) Heisenberg models in connection with the
high-$T_c$ cuprate superconductors. 
It has been argued by Haldane \cite{Haldane} and Chakravarty, Halperin and
Nelson \cite{Chakravarty} that the phase transition is 
described by the 2+1-dimensional non-linear O(3) $\sigma$ model predicting that
at approaching the transition point the 
spin-wave gap vanishes as $\Delta\sim (J_\perp-J_{\perp c})^\nu$ with  
critical index $\nu \approx 0.7$.
Assuming the $\sigma$-model description the universal dynamic 
and static properties of two-dimensional antiferromagnets 
in the vicinity of a zero-temperature phase transition have been studied in 
detail by Sachdev, Ye and Chubukov \cite{Sachdev,Chubukov0}.
However mapping of the Heisenberg model to the $\sigma$-model
has been proven only for the one-layer square lattice antiferromagnet 
which itself does not exhibit a zero temperature quantum phase transition.

In the present work we investigate the two-layer Heisenberg antiferromagnet
which has a zero temperature quantum phase transition from a
disordered dimer phase to a collinear Neel phase as the dimerization decreases.
For this model the numerical results of Singh, Gelfand and Huse \cite{Singh} and Weihong
\cite{Weihong} obtained by series expansions, as well as
Monte Carlo calculations performed by Sandvick and Scalapino \cite{Sandvik} 
demonstrate reasonable consistency of the critical index with that in the 
$\sigma$-model.
At the same time Sandvik and Vekic \cite{Sandvik1} using
quantum Monte-Carlo simulations have shown that for some other 2D 
antiferromagnetic models (dimers arranged in ladders and dimers in 
a staggered pattern) the critical exponents are 
different from the nonlinear $\sigma$-model prediction.
In this situation it is very important to analyze the behavior near the
critical points by an independent analytical method. Such a method,
based on the Brueckner perturbation theory, has been developed in
the paper \cite{Sushkov}. In the present work we apply it for the
calculation of the spin-wave gap critical index in the two-layer Heisenberg 
antiferromagnet. This method  gives an independent calculation
of the index for this model, and additionally it allows 
us to formulate the conditions when one could expect deviations from
the $\sigma$-model behavior.

We consider the two-layer, $S=1/2$, Heisenberg antiferromagnet 
described by 
\begin{equation}
\label{H}
H=J\sum_{<i,j>}({\bf S}_{1i}{\bf S}_{1j}+{\bf S}_{2i}{\bf S}_{2j})
+J_{\perp}\sum_{i}{\bf S}_{1i}{\bf S}_{2i}.
\end{equation}
The spins ${\bf S_1}$ and ${\bf S_2}$ represent two planes coupled
by $J_{\perp}$. Both couplings are antiferromagnetic 
($J,J_{\perp}>0$) and the first sum runs over nearest neighbors on a square lattice.
Considering the ``J-terms'' in (\ref{H}) as a perturbation one can introduce
the bond operator representation \cite{Chub,Sach} and {\it exactly}
map the Hamiltonian (\ref{H}) to the effective Hamiltonian \cite{Sushkov}:
\begin{equation}
\label{ham}
H_{eff}=H_2+H_4+H_U,
\end{equation}
\begin{equation}
H_2=\sum_{\bf k,\alpha}A_{\bf k}t^{\dag}_{\bf k,\alpha}t_{\bf k,\alpha}+
\frac{B_{\bf k}}{2}(t^{\dag}_{\bf k,\alpha},t^{\dag}_{-\bf k,\alpha}+h.c),
\end{equation}
\begin{equation}
\label{h4}
H_4=\frac{J}{2}\sum_{<i,j>,\alpha \beta}\{t_{\alpha i}^{\dag}t^{\dag}_{\beta j}
t_{\beta i}t_{\alpha j}-t^{\dag}_{\alpha i}t^{\dag}_{\alpha j}
t_{\beta i}t_{\beta j}\},
\end{equation}
\begin{equation}
\label{cst}
H_U=U\sum_{i,\alpha \beta}t^{\dag}_{\alpha i}t^{\dag}_{\beta i}t_{\beta i}
t_{\alpha i},\hspace{0.2cm} U\rightarrow \infty
,\end{equation}
where $t^{\dag}_{\alpha i}$ is the creation operator of the triplet
at the bond $i$ and $\alpha =x,y,z$ is the polarization of the triplet.
The operator $t^{\dag}_{{\bf k},\alpha}$ is the Fourier transform of  
$t^{\dag}_{\alpha i}$: $t^{\dag}_{{\bf k},\alpha}=
\frac{1}{\sqrt{N}}\sum_{\bf r}t_{{\bf r},\alpha}e^{i\bf {(k+k_0) r}}$.
As usual the momentum takes values inside the Brillouin zone
$-\pi< k_x \le \pi$, $-\pi< k_x \le \pi$, but we shift the argument
in the Fourier transform by ${\bf k_0}=(\pi,\pi)$.
In this notation the minimum of the spin-wave dispersion is at ${\bf k}=0$.
The coefficients are of the form 
$A_{\bf k}=J_{\perp}+2J\xi_{\bf k}$, $B_{\bf k}=2J\xi_{\bf k}$,
 $\xi_{\bf k}=-\frac{1}{2}(\mbox{cos}k_x+\mbox{cos}k_y)$.
An infinite on-site repulsion between triplets $H_U$ is introduced
to take into account the hard-core constraint 
$t_{\alpha i}^{\dag}t_{\beta i}^{\dag}=0$ 
(only one triplet can be excited on a bond).
The interaction $H_U$ gives the dominant contribution to the renormalization of 
the spin-wave spectrum. It has been demonstrated \cite{Sushkov} that 
in the Brueckner approximation this renormalization is described by 
the self-energy operator
\begin{equation}
\label{en}
\Sigma({\bf k},\omega)=\frac{4}{N}\sum_{\bf q}Z_{\bf q}v_{\bf q}^2
\Gamma({\bf k+q},\omega-\omega_{\bf q}),         
\end{equation}
where the scattering amplitude is
\begin{equation}
\label{ver}
\Gamma({\bf q},\omega)=
-\left(\frac{1}{N}\sum_{\bf p}\frac{Z_{\bf p}Z_{\bf q-p}
u_{\bf p}^2u_{\bf{q-p}}^2}
{\omega-\omega_{\bf p}-\omega_{\bf{q-p}}}\right)^{-1},
\end{equation}
and the spin wave spectrum $\omega_{\bf k}$, quasiparticle residue
$Z_{\bf k}$ and Bogoliubov parameters $u_{\bf k}$, $v_{\bf k}$
are given by the formulas:
\begin{equation}
\label{sp}
\omega_{\bf k}=Z_{\bf k}\sqrt{\tilde A_{\bf k}^2-\tilde B_{\bf k}^2},
\end{equation}
\begin{eqnarray}
\label{ab}
&&\tilde A_{\bf k}=J_{\perp}+2J\xi_{\bf k}+\Sigma({\bf k},0)+
4J\xi_{\bf k}\sum_{\bf q}\xi_{\bf q}Z_{\bf q}v_{\bf q}^2,\\
&&\tilde B_{\bf k}=2J\xi_{\bf k}-4J\xi_{\bf k}
\sum_{\bf q} \xi_{\bf q}Z_{\bf q} u_{\bf q}v_{\bf q},\nonumber
\end{eqnarray}
\begin{eqnarray}
\label{zuv}
&&Z_{\bf k}=\left(1-\left.\frac{\partial \Sigma}{\partial \omega}
\right|_{\omega=0}\right)^{-1},\\
&&u^2_{\bf k},v^2_{\bf k}=\frac{Z_{\bf k}\tilde A_{\bf k}}
{2\omega_{\bf k}}\pm\frac{1}{2}.\nonumber
\end{eqnarray}
These equations take also into account the quartic interaction (\ref{h4})
in one-loop approximation.
In order to find spectrum, equations (\ref{en},\ref{ver},\ref{sp},\ref{ab},\ref{zuv})
have to be solved self-consistently for $\Sigma({\bf k},0)$ and $Z_{\bf k}$.
The plot of the spin-wave gap
$\Delta$ versus $J_{\perp}/J$ is presented
in Fig.1. In the same figure we present results from the papers
\cite{Weihong,Sushkov} obtained by the dimer series expansions.

Close to the critical point ($\Delta \ll J$) and for small momenta
($k\ll 1$) the dispersion can be represented as
\begin{equation}
\label{diss}
\omega_{\bf k} \approx \sqrt{\Delta^2+c^2k^2},
\end{equation}
where $c=1.85J$ is the spin-wave velocity \cite{Weihong,Sushkov,Gelfand}.
To find the critical index let us write eq. (\ref{sp}) at the point
${\bf k}=0$
\begin{equation}
\label{d0}
\Delta^2=Z_0^2(\tilde A_0^2-\tilde B_0^2).
\end{equation}
It is convenient to introduce the values of $\tilde A_0$ and $\tilde B_0$
at the critical point: $\tilde A_{0c}=-\tilde B_{0c}$.
Let us vary $J_{\perp}$ keeping $J$ fixed, and let us introduce the deviation
from the critical point $\delta J_{\perp}=J_{\perp}-J_{\perp c}$.
Considering $\delta J_{\perp}$ and $\Delta$ as independent variables
and using eqs. (\ref{ab},\ref{zuv},\ref{diss}) we find the variations of
$\tilde A_0$ and $\tilde B_0$:
\begin{eqnarray}
\label{abv}
&&\tilde A_0=\tilde A_{0c}+\delta J_{\perp}+\delta \Sigma (0,0)-
{{Z_0 J \tilde A_{0c}}\over{\pi c^2}}\Delta, \\
&&\tilde B_0=\tilde B_{0c}+
{{Z_0 J \tilde A_{0c}}\over{\pi c^2}}\Delta. \nonumber
\end{eqnarray}
To find the variation of the self-energy notice that according to eq. (\ref{ver})
$\Gamma({\bf q},-\omega_{\bf q})\propto q$ at $\Delta/c \sim q \ll 1$.
Therefore from eq. (\ref{en}) we find
\begin{equation}
\label{ds}
\delta\Sigma(0,0) \propto \sum_{\bf q}q \delta v_{\bf q}^2 \propto
\Delta^2\ln{J\over{\Delta}}.
\end{equation}
Terms linear in $\Delta$ are canceled out after substitution of (\ref{abv})
and (\ref{ds}) into eq. (\ref{d0}), and neglecting logarithmic dependence
we find
\begin{equation}
\label{nbr}
\Delta \propto \sqrt{\delta J_{\perp}}.
\end{equation}
Thus the critical index in the Brueckner approximation is $\nu=0.5$.
It is known from nuclear and atomic physics\cite{Dzuba}
that the Brueckner approximation usually works pretty well even in 
systems with high density, but parametrically it is justified only
at the low density limit. In essence it is the dilute gas approximation. It has been
demonstrated \cite{Sushkov} that for the model under consideration
the actual small parameter is $n_b\ln(J/\Delta)$, where $n_b\approx 0.1$
is density of the triplet excitations. It is clear that when the gap is very 
small this parameter is becoming large and the gas approximation can fail.
To analyze the situation numerically we plot in the Fig.2 the
functions
\begin{eqnarray}
\label{f12}
&&f_1(x)=\left({{d\ln(\Delta/J)}\over{dx}}\right)^{-1},\\ 
&&f_2(x)=-\left({{d\ln(\Delta/J)}\over{d x}}\right)^2\left/
{{d^2\ln(\Delta/J)}\over{d^2x}}\right.,\nonumber
\end{eqnarray}
where $x=J_{\perp}/J$ and $\Delta$ is found
from eqs. (\ref{en},\ref{ver},\ref{sp},\ref{ab},\ref{zuv}).
For pure scaling behavior $\left[\Delta \propto (\delta J_{\perp})^{\nu}
\right]$ the function $f_1(x)$ is linear  and the
function $f_2$ gives the index: $f_2(x)=\nu$.
Keeping in mind that when the gap is small  ($\ln(J/\Delta) \gg 1$)
our approximation is not justified, we conclude from the plot of $f_2$
that the estimate of the critical index is $\nu \in [0.6,0.75]$ or
\begin{equation}
\label{nu}
\nu \approx 0.67\pm0.07.
\end{equation}

Another way to look at the critical index is to expand the Brueckner
equations in powers of the density $n_b$. In the leading approximation
$Z_{\bf q}=u_{\bf q}=1$ and therefore the vertex (\ref{ver}) takes
the form
\begin{equation}
\label{ver1}
 \Gamma({\bf q},-\omega_{\bf q})=\left(
\int\frac{d^2{\bf p}}{(2\pi)^2}\frac{1}{\omega_{\bf q}
+\omega_{\bf p}+\omega_{\bf {p-q}}}\right)^{-1}.
\end{equation}
In contrast to (\ref{ver}) it does not vanish at $q \to 0$.
Let us denote $\Gamma(0,0)=\Gamma_c$. 
The last terms in the equations (\ref{abv}) are due to the quartic
interaction $H_4$ (\ref{h4}). This interaction is relatively small
and therefore we neglect these terms. Below we prove that they are
really small. So instead of eqs. (\ref{abv}) we have
\begin{eqnarray}
\label{abv1}
&&\tilde A_0=\tilde A_{0c}+\delta J_{\perp}+\delta \Sigma (0,0) \\
&&\tilde B_0=\tilde B_{0c}. \nonumber
\end{eqnarray}
If we substitute this into eq.(\ref{d0}) and neglect terms quadratic in
$\Delta$  we find that variation of $\tilde A_0$ must vanish
\begin{equation}
\label{varA}
\delta \tilde A_0=\delta J_{\perp}+\delta \Sigma (0,0)=0.
\end{equation}
The variation of the self-energy should be found from eq. (\ref{en})
 \begin{equation}
\label{crit}
\delta\Sigma(0,0)=4\int\frac{d^2{\bf q}}{(2\pi)^2}
\Gamma({\bf q},-\omega_{\bf q})\delta v_{\bf q}^2+
4\int\frac{d^2{\bf q}}{(2\pi)^2}\delta\Gamma({\bf q}
,-\omega_{\bf q}) v_{\bf q}^2.
\end{equation}
The main contribution to the first integral in this formula comes from small momenta
($q\sim \Delta/c \ll 1$) since 
\begin{equation}
\delta v_{\bf q}^2 ={1\over2}\left\{
{{\delta \tilde A_{\bf q}}\over{\omega_{\bf q}}}+\tilde A_{\bf q}\delta\left(
{1\over{\omega_{\bf q}}}\right)\right\}\approx
-{{\tilde A_{0c}\Delta^2}\over{4(\Delta^2+c^2q^2)^{3/2}}}.
\end{equation}
The variation of $\tilde A$ in this formula vanishes according to 
eq. (\ref{varA}). Then the integral can be easily evaluated and the result is
$-\Gamma_c\tilde A_{0c}\Delta/\pi c^2$. The main contribution to the second integral in (\ref{crit})
comes from large momenta $(q \sim 1)$ where we can write
$\delta \Gamma({\bf q},-\omega_{\bf q}) = \Gamma^{\prime}\delta J_{\perp}$.
It is obvious from (\ref{ver1}) that at $J_{\perp} \gg J$ the derivative 
$\Gamma^{\prime}=3$. However we need this derivative near the critical
point where the numerical calculation shows that $\Gamma^{\prime} \approx 2.9$ 
independent of momenta. Altogether the variation (\ref{crit}) can be
represented as
\begin{equation}
\label{vert1}
\delta\Sigma(0,0)\approx -\frac{\Gamma_c \tilde A_{0c}}{\pi c^2}\Delta
+\frac{4}{3}\Gamma'n_b\delta J_{\perp},
\end{equation}
where $n_b=3\sum_{\bf q}v_{\bf q}^2\approx 0.12$ is the density of spin-wave
excitations at the critical point. 
At this step we can check how small is the neglected quartic interaction term 
in eqs. (\ref{abv}).  The ratio of this term to the ``$\Delta$-term''
in (\ref{vert1}) is $J/\Gamma_c \approx 0.15$, since $\Gamma_c\approx 6.3J$.
After substitution of (\ref{vert1}) into eq. (\ref{varA})
we find the relation between the gap and $\delta J_{\perp}$
\begin{equation}
\label{gap1}
\Delta\approx\frac{\pi c^2}{\Gamma_c A_{0c}}(1+\frac{4}{3}\Gamma'n_b)
(J_{\perp}-J_{\perp c})\approx 1.1(J_{\perp}-J_{\perp c}).
\end{equation}
The coefficient 1.1 corresponds to
$\Gamma_c\approx 6.3J$, $A_{0c}\approx 2.4J$, 
$c\approx 1.9J$, $\Gamma'\approx 2.9$, and $n_b\approx 0.12$ which
have been found by numerical solution of the linearized in
density eqs. (\ref{en},\ref{sp},\ref{ab},\ref{zuv}).
We remind that the linearization means that the vertex is taken
from eq. (\ref{ver1}) and the residues in (\ref{ab}) are replaced
by unity ($Z_{\bf q}=1$). The numerical solution also gives 
$\Delta(J_{\perp})$ shown in Fig.1. The slope at $J_{\perp}=J_{\perp c}$ is 
in perfect agreement with the semianalytical formula (\ref{gap1}).
Thus the leading term in powers of the density gives the critical exponent $\nu=1$,
the same as  the mean field approximation.

Now, consider the first correction due to the triplet density $n_b$.
We will keep only those terms which contain additional $\ln q$ or $\ln(\Delta/J)$,
so the parameter is $n_b \ln(\Delta/J)$. The $n_b \ln(\Delta/J)$ terms
arise only from expansion of the vertex (\ref{ver}), replacing
$u_{\bf p}^2 u_{\bf q-p}^2=(1+v_{\bf p}^2)(1+v_{\bf q-p}^2)\approx
1+v_{\bf p}^2+v_{\bf q-p}^2$ we obtain
\begin{equation}
\label{ver2}
\Gamma({\bf q},-\omega_{\bf q})=
\left(\int\frac{d^2p}{(2\pi)^2}
\frac{1+v_{\bf p}^2+v_{\bf {q-p}}^2}
{\omega_{\bf q}+\omega_{\bf p}+\omega_{\bf {q-p}}}\right)^{-1}.
\end{equation}
After simple integration
one can find that for small $q$ ($q\sim \Delta/c \ll 1$) the vertex is
\begin{equation}
\Gamma(q,-\omega_q)\approx\Gamma_c+\frac{\Gamma_c^2\tilde A}
{4\pi c^2}\mbox{ln}q.
\end{equation}
Substitution into (\ref{crit}) gives the variation of the self-energy
\begin{equation}
\delta\Sigma(0,0)\approx -\frac{\Gamma_c\tilde A_{0c}}{\pi c^2}\Delta
\left(1+\frac{\Gamma_c\tilde A_{0c}}{4\pi c^2}\mbox{ln}\frac{\Delta}{J}\right)
+\frac{4}{3}\Gamma'n_b\delta J_\perp,
\end{equation}
which together with eq. (\ref{varA}) results in
\begin{equation}
\label{log}
\Delta=\frac{\pi c^2}{\Gamma_c \tilde A_{0c}}
(1+\frac{4}{3}\Gamma'n_b)\delta J_{\perp}\left(1-
\frac{\Gamma_c \tilde A_{0c}}{4\pi c^2}\ln\frac{\delta J_{\perp}}{J}\right).
\end{equation}
Let us assume the scaling behavior $\Delta \propto (\delta J_{\perp})^{\nu}$
with $\nu=1-\beta$.  Expanding this formula in powers of $\beta$ and 
comparing with (\ref{log}) we find
\begin{equation}
\nu=1-\frac{\Gamma_c \tilde A_{0c}}{4\pi c^2}\approx 0.67,
\end{equation}
which agrees with the estimate (\ref{nu}) and with the result of the
$\sigma$-model approach($\nu \approx 0.70$ \cite{Ferer}).

In conclusion, using the Bruekner approach for two-layer Heisenberg
antiferromagnet we have calculated the spin-wave critical index
for the zero temperature quantum phase transition. The result is in
reasonable agreement with that of the nonlinear O(3) $\sigma$-model.
This agreement is due to the relative smallness of the quartic
interaction for the  model under consideration. In this situation
the hard-core constraint is the most important and it is very natural that the
result is similar to that of the $\sigma$-model. However this situation is not general. There are many models where the quartic interaction
is very important. It can even produce bound states of triplet
spin waves \cite{bs} which effectively change the number of relevant
degrees of freedom. In this situation one can expect a very substantial
deviation from the simple $\sigma$-model. An important example of such
a system is the 2D $J_1-J_2$ model where the singlet bound state has
an extremely low energy \cite{K}.

We are very grateful to V. N. Kotov, R. R. P. Singh and Z. Weihong for 
stimulating  discussions. This work was supported by a grant from the 
Australian Research Council.

FIGURE CAPTIONS

Fig.1 Triplet gap $\Delta/J$ as a function of  $x=J_\perp/J$.
Solid and dashed lines are the results of the self-consistent
solution using the Bruekner approach and the Bruekner equations linearized in 
density respectively. The dots (with error bars) are estimates obtained by dimer series expansions.

Fig.2 Functions $f_1(x)=\left[\frac{d\ln(\Delta/J)}{dx}\right]^{-1}$ and
$f_2(x)=-\left[\frac{d\ln(\Delta/J)}{dx}\right]^2/\frac{d^2\ln(\Delta/J)}{dx^2}
$ versus $x=J_\perp/J$ (see text).
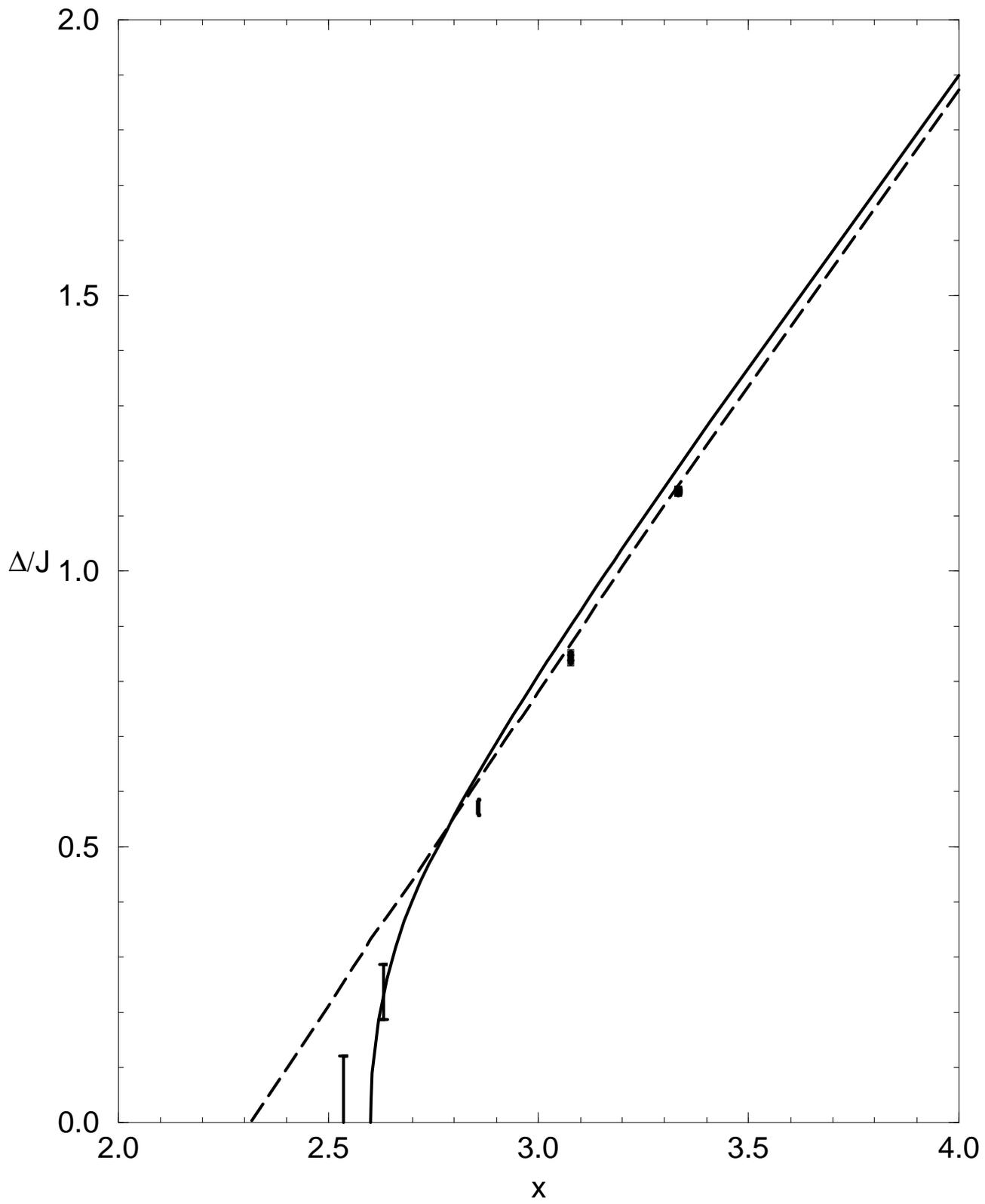
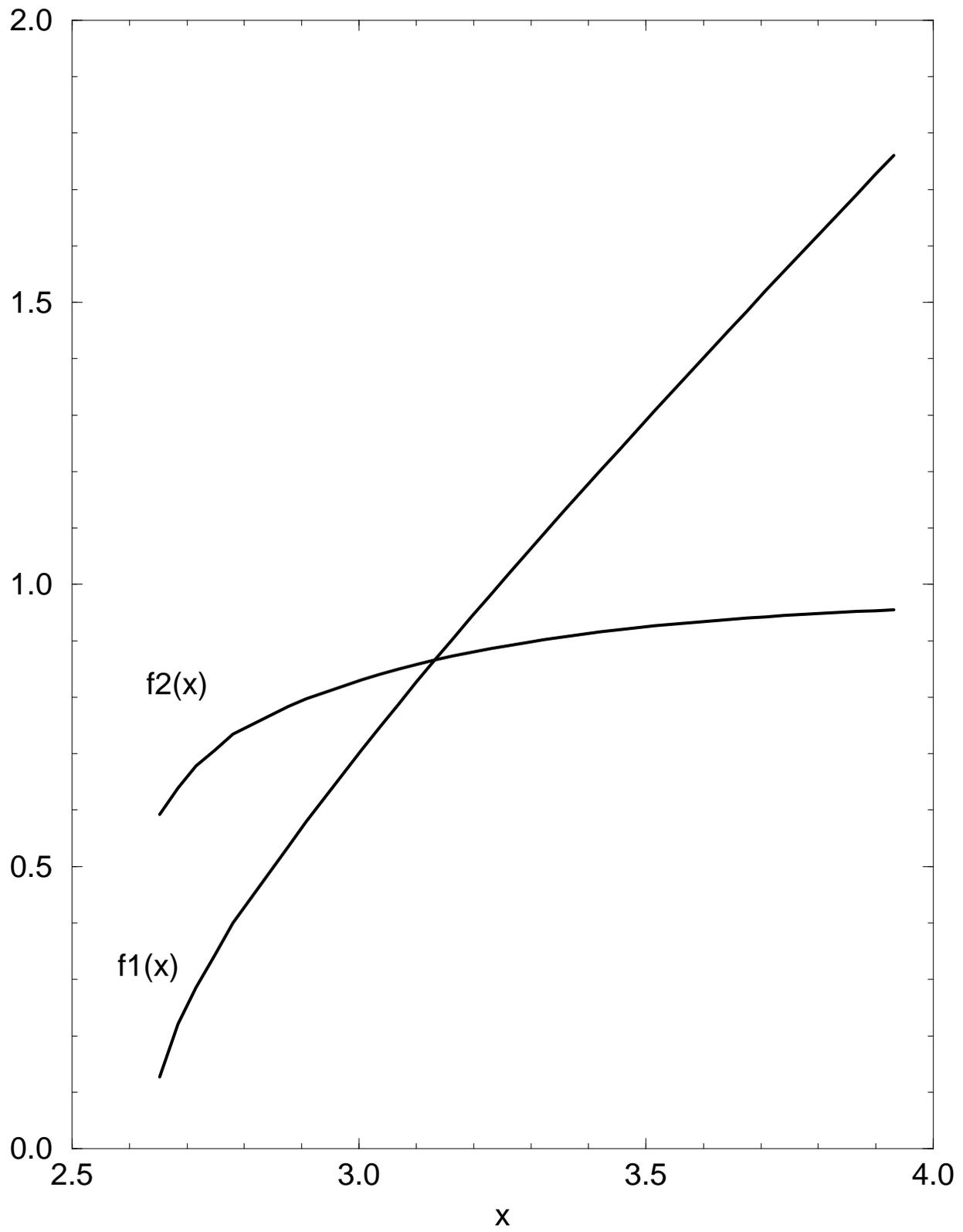
\end{document}